\documentclass[a4paper,fleqn,usenatbib]{mnras}
\pdfoutput=1
\usepackage[T1]{fontenc}
\usepackage{ae,aecompl}

\usepackage{graphicx}	% Including figure files
\usepackage{amsmath}	% Advanced maths commands
\usepackage{amssymb}	% Extra maths symbols

\title[{Limits on the validity of the} thin-layer {model}]{Limits on the validity of the thin-layer {model of} the ionosphere for radio interferometric calibration}

\author[P. L. Martin et al.]{
Poppy L. Martin,$^{1,2}$\thanks{E-mail: poppy.martin@postgrad.manchester.ac.uk}
Justin D. Bray,$^{2}$
and Anna M. M. Scaife$^{2}$
\\
$^{1}$Department of Physics \& Astronomy, University of Southampton, Southampton, SO17 1BJ, UK\\
$^{2}$JBCA, School of Physics \& Astronomy, University of Manchester, Oxford Road, Manchester, M13 9PL, UK\\
}

\date{Accepted XXX. Received YYY; in original form ZZZ}

\pubyear{2015}

\begin{document}
\label{firstpage}
\pagerange{\pageref{firstpage}--\pageref{lastpage}}
\maketitle

\begin{abstract}
For a ground-based radio interferometer observing at low frequencies, the ionosphere causes propagation delays and refraction of cosmic radio waves which result in phase errors in the received signal. These phase errors {can be} corrected using a calibration method that assumes a two-dimensional phase screen at a fixed altitude above the surface of the Earth, known as the thin-layer {model}. Here we investigate the validity of the thin-layer {model} and provide a simple equation with which users can check when this approximation can be applied to observations for varying time {of day}, zenith angle, {interferometer latitude}, baseline {length}, {ionospheric electron content} and observing frequency. 
% TODO: once the rest of the manuscript is settled, consider revising the abstract
\end{abstract}


%%%%%%%%%%%%%%%%%%%%%%%%%%%%%%%%%%%%%%%%%%%%%%%%%%

%%%%%%%%%%%%%%%%% BODY OF PAPER %%%%%%%%%%%%%%%%%%

\section{Introduction}
\label{sec:intro}

{Radio waves from astronomical sources, particularly at low frequencies, experience phase delays as they propagate through the low-density plasma of the Earth's ionosphere.  This effect is of concern in radio interferometry because it affects measurements of the phase differences between pairs of antennas that are used by radio {synthesis} telescopes to reconstruct images of radio emission from the sky.}  Whilst telescopes operating at higher radio frequencies (\textgreater 1 GHz) {are} generally able to {correct for ionospheric effects using direction-independent, time-varying gain phases from self-calibration}, arrays operating at lower frequencies such as the {{Karl G. Jansky Very Large Array (VLA;} 58--470~MHz; \citealt{perley11})}, { the Giant Metrewave Radio Telescope (GMRT;} 153--610~MHz; \citealt{swarup91}), {the Murchison Widefield Array (MWA;} 80--300~MHz; \citealt{tingay2012}) and {the Low-Frequency Array (LOFAR;} 30--240~MHz; \citealt{vanhaarlem13}) have found that {direction-dependent ionospheric effects} can {limit} the achievable dynamic range of astronomical observations. {This will also affect future low frequency arrays such as { the Square Kilometre Array (SKA-low; 50--350~MHz; \citealt{dewdney13}).}}

\begin{figure}
\begin{center}
\includegraphics[width=\columnwidth]{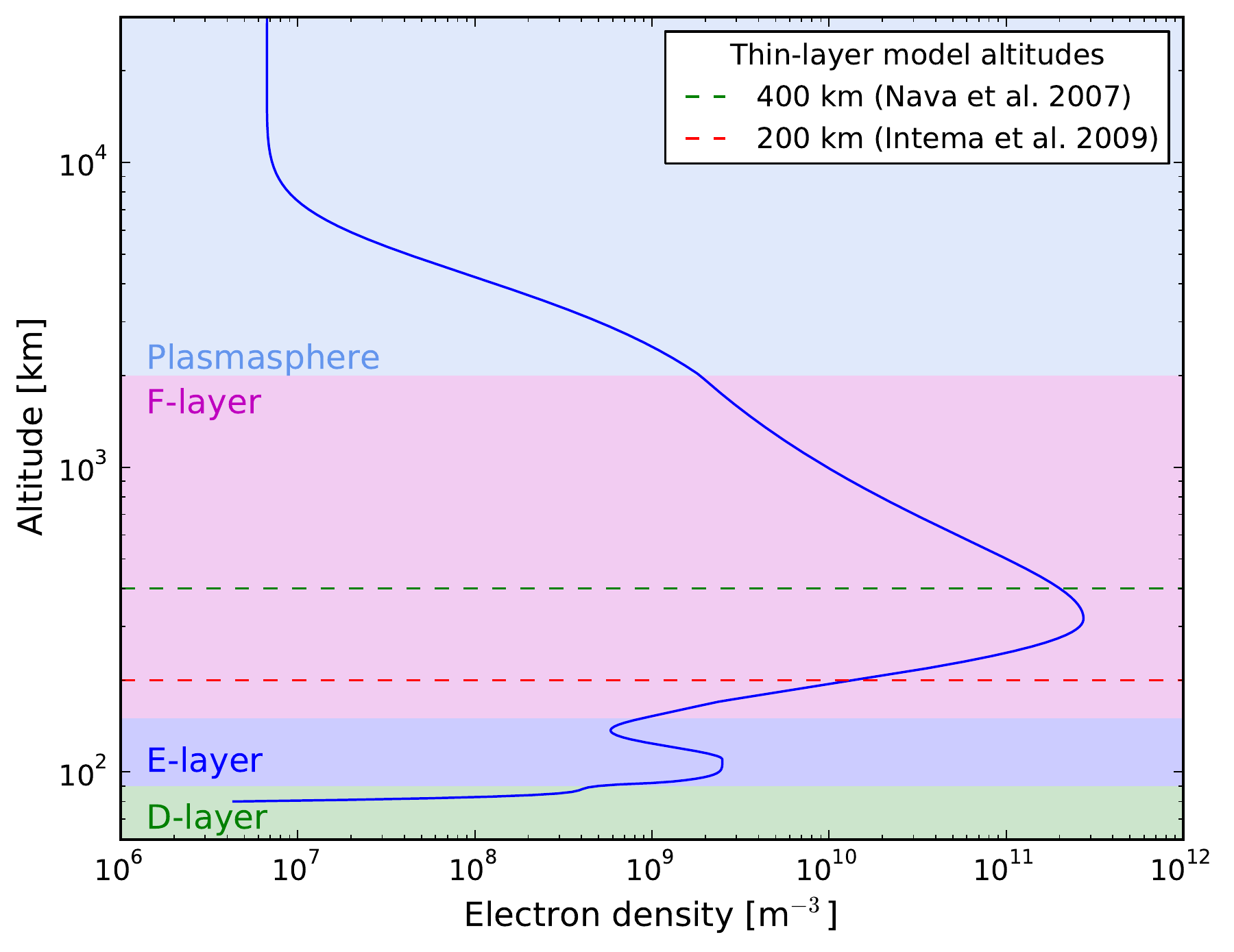}
\caption{{Typical} profile of {ionospheric electron density, showing layers of the ionosphere and some altitudes assumed in applications of the thin-layer model.  Electron density values below} 2000~km {were} obtained from {the} IRI { for 03:00~local time on 17~February 2015 at 40$^{\circ}$ geomagnetic latitude and 0$^{\circ}$ geomagnetic longitude}, {while values above this altitude were obtained from the plasmaspheric model of} \citet{gallagher88} {for the same geomagnetic latitude and time of day} {and scaled to fit (see text)}.}
\label{fig:heightprof}
\end{center}
\end{figure}

{Data reduction in radio interferometric imaging compensates for this effect through ionospheric calibration.  This involves developing a model of the spatial and temporal variation of the ionospheric electron content, calculating the resulting effect on the phases measured by the telescope, and {applying the direction-dependent phase corrections to the data \citep{intema09}}.  Because the ionospheric electron content is concentrated at a narrow range of altitudes (see Figure~\ref{fig:heightprof}), it} is typically modelled as a spherical shell of infinitesimal thickness, centred on the Earth.  The {assumed} altitude {of} this thin layer above the Earth's surface varies{, ranging from 200~km to 500~km \citep{nava07,cohen09,intema09,intemacode14}.}

{The thin-layer model is not a perfect representation of the ionosphere, and under certain conditions --- a telescope operating with widely-separated antennas at low frequencies, or observations with a highly active ionosphere --- the error associated with this approximation may limit the dynamic range of the resulting radio image.}  \citet{intema09} state that it is unclear under {what} conditions {this occurs.  More complex models are possible to avoid this error --- for example, \citet{intema11} model the ionosphere as three distinct layers at 100~km, 200~km and 400~km --- but it would be helpful to have a simple test to determine when such an approach is required.}

In this paper, we investigate the range of parameters over which the thin-layer {model} is valid. Section~\ref{sec:thinlayermodel} introduces the thin-layer model, {Section~\ref{sec:ionocal} describes its application to ionospheric calibration in radio interferometry, }and Section~\ref{sec:testingthinlayer} {discusses the error that results from the thin-layer model, and the consequent error in the ionospheric phase calibration of a radio synthesis telescope.  We present our final result in the form of an} equation with which radio {astronomers} can check when {the thin-layer} {model} can be {safely} applied to observations{, and briefly summarise in Section~\ref{sec:conc}}.

%----------------------------------------------------------------------------------------------------------------------------------------------------------

\section{The thin-layer model}
\label{sec:thinlayermodel}

\begin{figure}
\begin{center}
\includegraphics[width=\columnwidth]{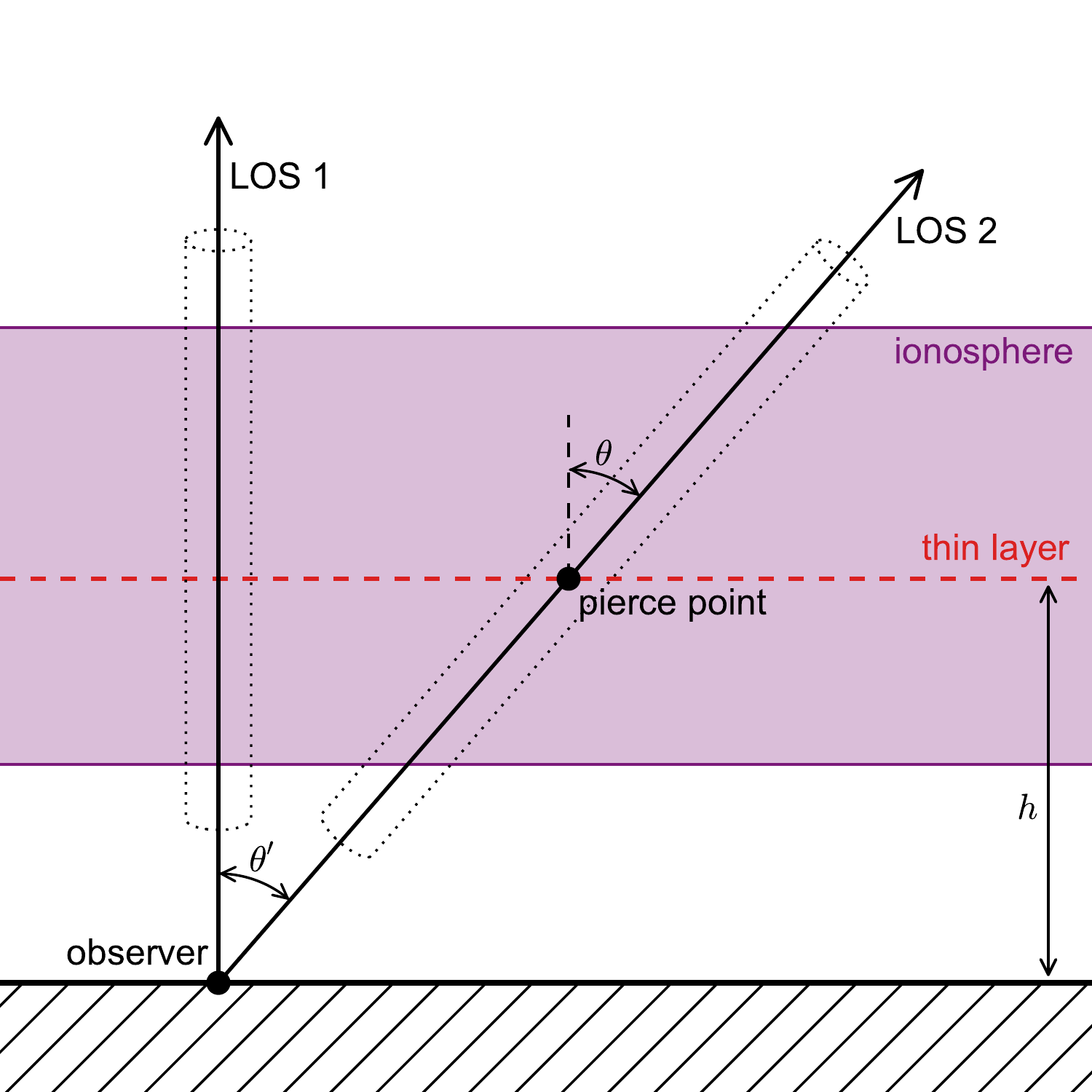}
\caption{{Illustration of vertical and slanted lines of sight through the ionosphere.  The column density of electrons along a vertical line of sight (`LOS~1') is the VTEC.  The column density of electrons along a slanted line of sight (`LOS~2') is the STEC, and is related to the VTEC in the thin-layer {model} by the angle $\theta$ at which it intersects the assumed thin layer, per Equation~\ref{eq:VTEC}.  In this simplified case, ignoring the curvature of the Earth, $\theta$ is independent of the altitude $h$ of the thin layer, and equal to the zenith angle $\theta'$ of the line of sight from the observer.}}
\label{fig:justinSTEC}
\end{center}
\end{figure}

{The quantity of interest for determining the ionospheric phase delay is the electron column density} or total electron content (TEC){, commonly} measured in TEC units (TECU), where one TECU is $10^{16}$~electrons\,m$^{-2}$.  {In the thin-layer model,} the ionosphere {is represented} as a thin spherical shell surrounding the Earth {at a fixed altitude $h$ above the mean Earth surface.  The true distribution of ionospheric electron content with altitude shown in Figure~\ref{fig:heightprof} is neglected, and instead described only by the equivalent TEC along a vertical column (vertical TEC; VTEC) at each point on the shell. The} slant TEC (STEC) {along a line of sight can then be calculated based on the VTEC at} the point at which the {line of sight pierces the thin layer, as}
\begin{equation}
{\rm STEC} = \frac{\rm VTEC}{\cos\theta} ,
\label{eq:VTEC}
\end{equation}
where $\theta$ is the angle {of} the {line of sight to the} thin layer {at the pierce point}.  {In the simplified flat-Earth case,} shown in Figure~\ref{fig:justinSTEC}{,} this is also the zenith angle {$\theta'$}of the {line of sight from the observer}.  {Neglecting horizontal variation of the TEC, the thin-layer {model} is in this case perfectly accurate, and not dependent on the value chosen for $h$.}

\begin{figure}
\begin{center}
\includegraphics[width=\columnwidth]{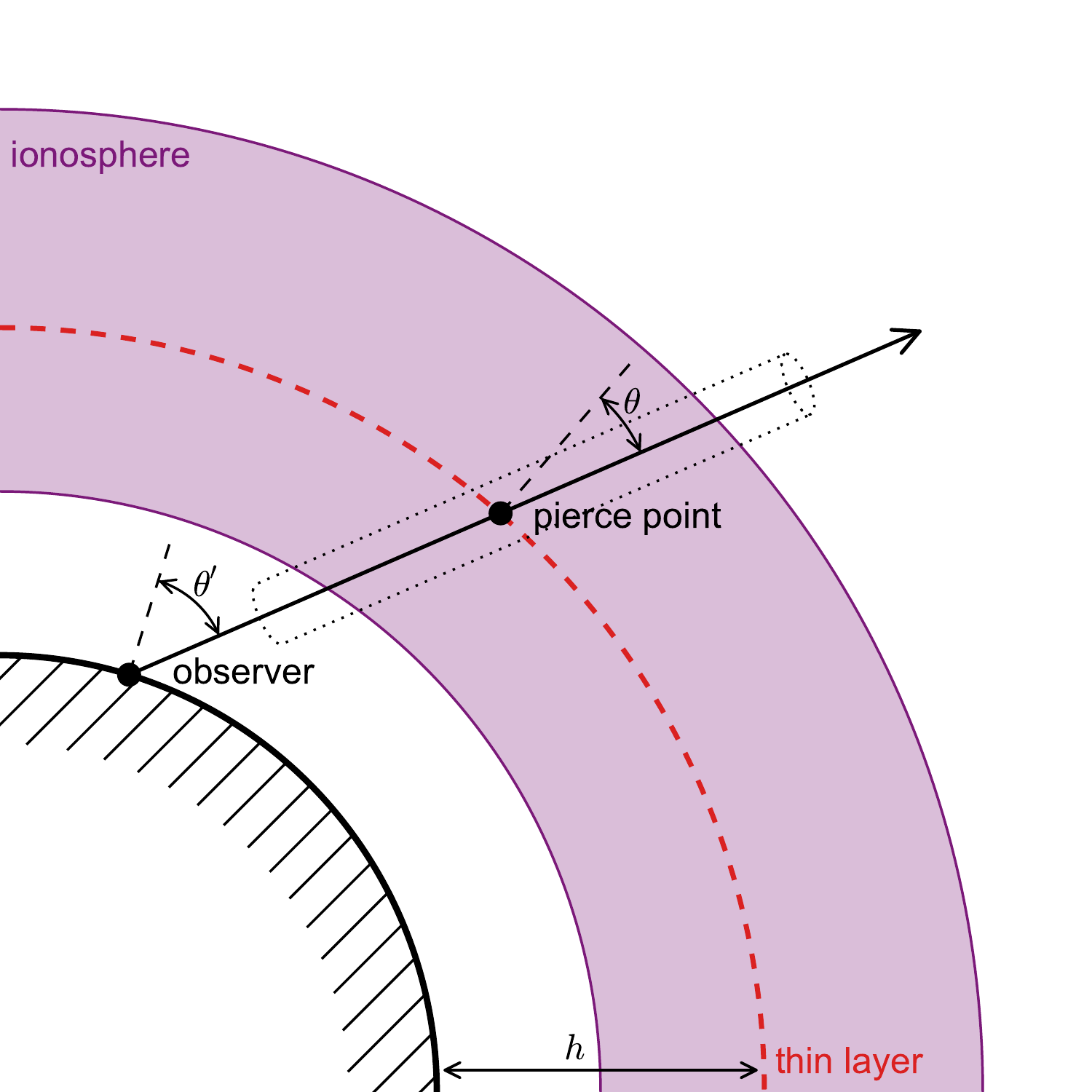}
\caption{{Illustration of a slanted line of sight through the ionosphere for a spherical Earth.  In the thin-layer {model}, the VTEC and STEC are related by the pierce angle $\theta$, as in Figure~\ref{fig:justinSTEC}.  However, unlike the flat-Earth case, the pierce angle $\theta$ is not equal to the zenith angle $\theta'$, and depends on the assumed altitude $h$ of the thin layer.  Since $h$ is only an approximation to the true distribution of the ionospheric electron content across a range of altitudes, this leads to inaccuracy in the thin-layer model.}}
\label{fig:justincurvedSTEC}
\end{center}
\end{figure}

{For} a more realistic, spherical Earth{, as shown in} Figure~\ref{fig:justincurvedSTEC}{, this is no longer true. { The angle of the line of sight to the thin layer at the pierce point, $\theta$, and the zenith angle of the line of sight of the observer, {$\theta'$} can be related by the law of sines \citep{thompson01}}}
\begin{equation}
\sin\theta = \frac{R_{\rm E}}{R_{\rm E} + h} \sin\theta'
\end{equation}
{where} \mbox{$R_{\rm E} = 6,371$}~km is the {mean} radius of the Earth {and hence, from Equation~\ref{eq:VTEC},}
\begin{equation}
{\rm STEC} = {\rm VTEC} \times \left(1-\left(\frac{R_{\rm E}}{R_{\rm E} + h}\sin\theta'\right)^2\right)^{-\frac{1}{2}} .
\label{eq:STEC}
\end{equation}
{In this case, even if horizontal variation of the TEC is neglected, the STEC calculated with the thin-layer model will differ from the true electron column density along the line of sight, depending on the assumed altitude $h$ of the thin layer.}

{To illustrate the imprecision of the thin-layer model due to the curvature of the Earth, we will compare STEC$_{\rm 2D}$, calculated for a thin-layer ionosphere at altitude $h$ per Equation~\ref{eq:STEC}, with STEC$_{\rm 3D}$, calculated for a three-dimensional ionosphere with a realistic altitude profile of electron density.  To calculate STEC$_{\rm 3D}$, we divide our profile into individual layers of 1~km thickness and approximate each as a distinct thin layer, taking}
\begin{equation}
{\rm STEC_{\rm 3D}} = \sum\limits_{i=60}^{{30,000}} {\rm STEC}_i
\label{eq:STECtot}
\end{equation}
where STEC$_i$ is the STEC calculated {per Equation~\ref{eq:STEC}} from the VTEC value {for an} individual layer {at an altitude \mbox{$h = i$}~km}.  We obtain our electron density profile by combining separate models for the ionosphere proper and for the plasmasphere, a high-altitude {component of the ionosphere which} typically {contributes} about 10\% to the {daytime} TEC \citep{yizengaw08}{.  For the ionosphere up to an altitude of 2,000~km we} use the International Reference Ionosphere (IRI; \citealt{bilitza11}){, a historical model based on contemporaneous measurements; we use the profile for} {03:00~UT} on 17~February 2015 { assuming a pierce point of the thin layer at {$40^\circ$} geomagnetic latitude and $0^\circ$ geomagnetic longitude}. {The IRI model only extends up to altitudes of 2000~km, therefore, f}{or the plasmasphere at higher altitudes, we use the model of} \citet{gallagher88}{,} a parameterised fit to historical variation with latitude and time of day. We renormalise the latter to match the former at the altitude where our profile transitions between the two models. The resulting profile is shown in Figure~\ref{fig:heightprof}.

\begin{figure}
\begin{center}
\includegraphics[width=\columnwidth]{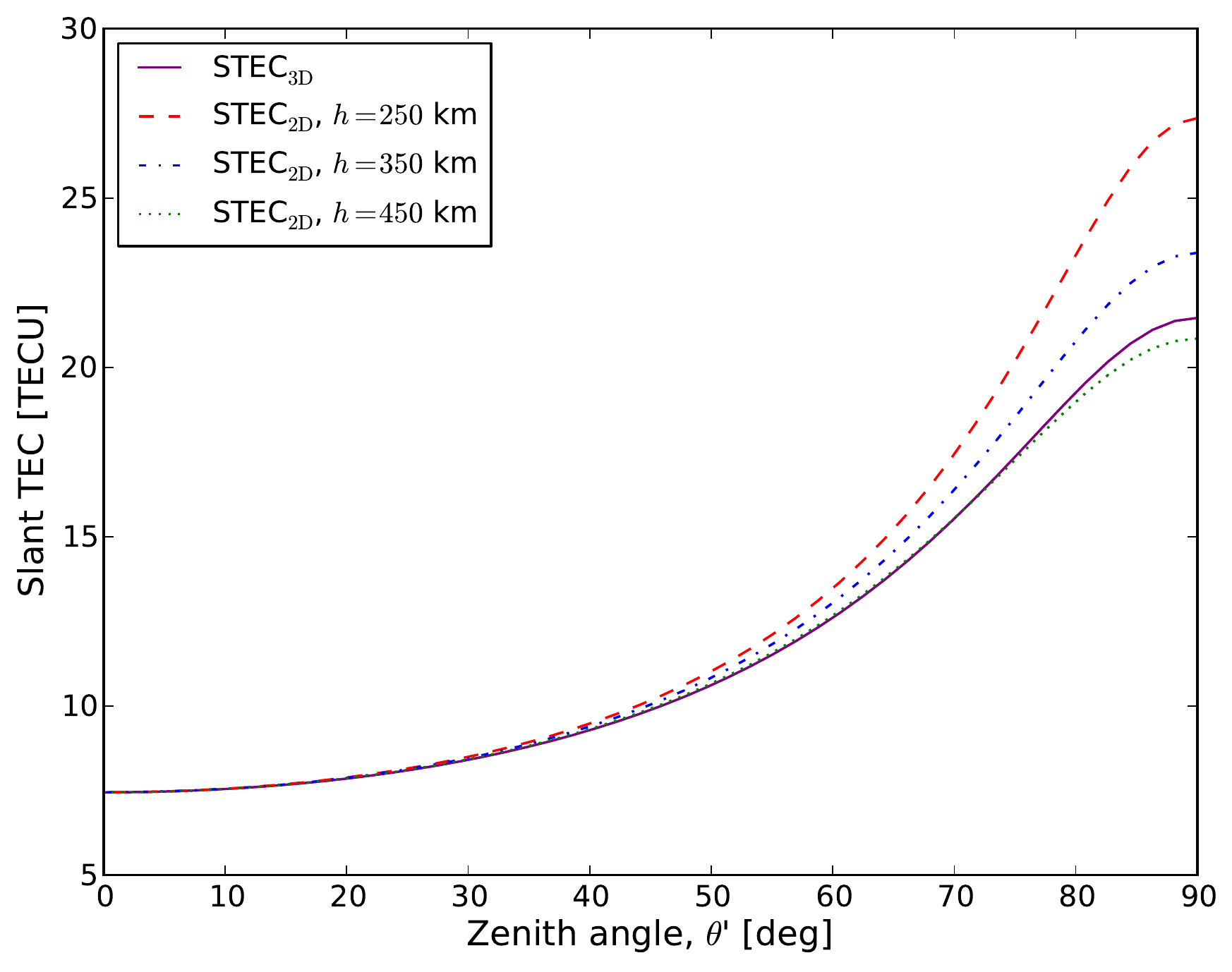}
\caption{{STEC$_{\rm 3D}$ calculated per Equation~\ref{eq:STECtot} for the electron density profile shown in Figure~\ref{fig:heightprof}, and STEC$_{\rm 2D}$ obtained for the same profile {using} the thin-layer {model} in Equation~\ref{eq:STEC}, both for a range of zenith angles $\theta'$ as shown in Figure~\ref{fig:justincurvedSTEC}.  All definitions of STEC are equivalent for a zenith angle of zero, at which they simply reproduce the VTEC, but generally diverge for lines of sight closer to the horizon.  The accuracy of the thin-layer {model} depends on the assumed thin-layer altitude $h$, with \mbox{$h = 450$}~km being approximately correct over the widest range of zenith angles.}}
\label{fig:SlantTECtopplot2}
\end{center}
\end{figure}

{The comparison between} STEC$_{\rm 2D}$ {calculated assuming a} thin-layer {model} and STEC$_{\rm 3D}$ {calculated for a realistic three-dimensional ionosphere is shown in} Figure~\ref{fig:SlantTECtopplot2}.  {The thin-layer model significantly overestimates the STEC at large zenith angles for} assumed thin-layer altitudes of {250~km and 350~km, but provides a somewhat closer approximation with an altitude of 450~km.  The applicability of the thin-layer model for phase calibration does not, however, depend directly on the precision with which it reproduces the STEC.  We will discuss this in more detail in Sections~\ref{sec:ionocal} and~\ref{sec:testingthinlayer}.}

%----------------------------------------------------------------------------------------------------------------------------------------------------------

\section{Ionospheric phase calibration}
\label{sec:ionocal}

{Radio interferometry depends upon measurements of the phase difference between pairs of antennas separated by baseline vectors.  If the lines of sight to a radio source from two antennas have different STECs, the ionosphere will introduce a different phase on each antenna, causing an error in the phase measurement.  The ionospheric phase delay is \citep{intema09}}
 \begin{align}
  \phi &\approx \frac{e^2}{4 \pi \varepsilon_0 m_{\rm e} c \nu} \, {\rm STEC} \\
   &\ {\approx 4840^\circ \times \left( \frac{\nu}{\rm 100~MHz} \right)^{-1} \left( \frac{\rm STEC}{\rm TECU} \right) }
   \label{eq:phiion}
 \end{align}
{where $e$ is the electron charge, $\varepsilon_0$ is the permittivity of free space, $m_{\rm e}$ is the mass of an electron, $c$ is the speed of light in a vacuum and $\nu$ is the radio frequency.  If the distribution of ionospheric electron content can be determined, the difference in STEC between two lines of sight can be calculated, and the resulting phase error subtracted from the measurements.}

\begin{figure}
\begin{center}
\includegraphics[width=\columnwidth]{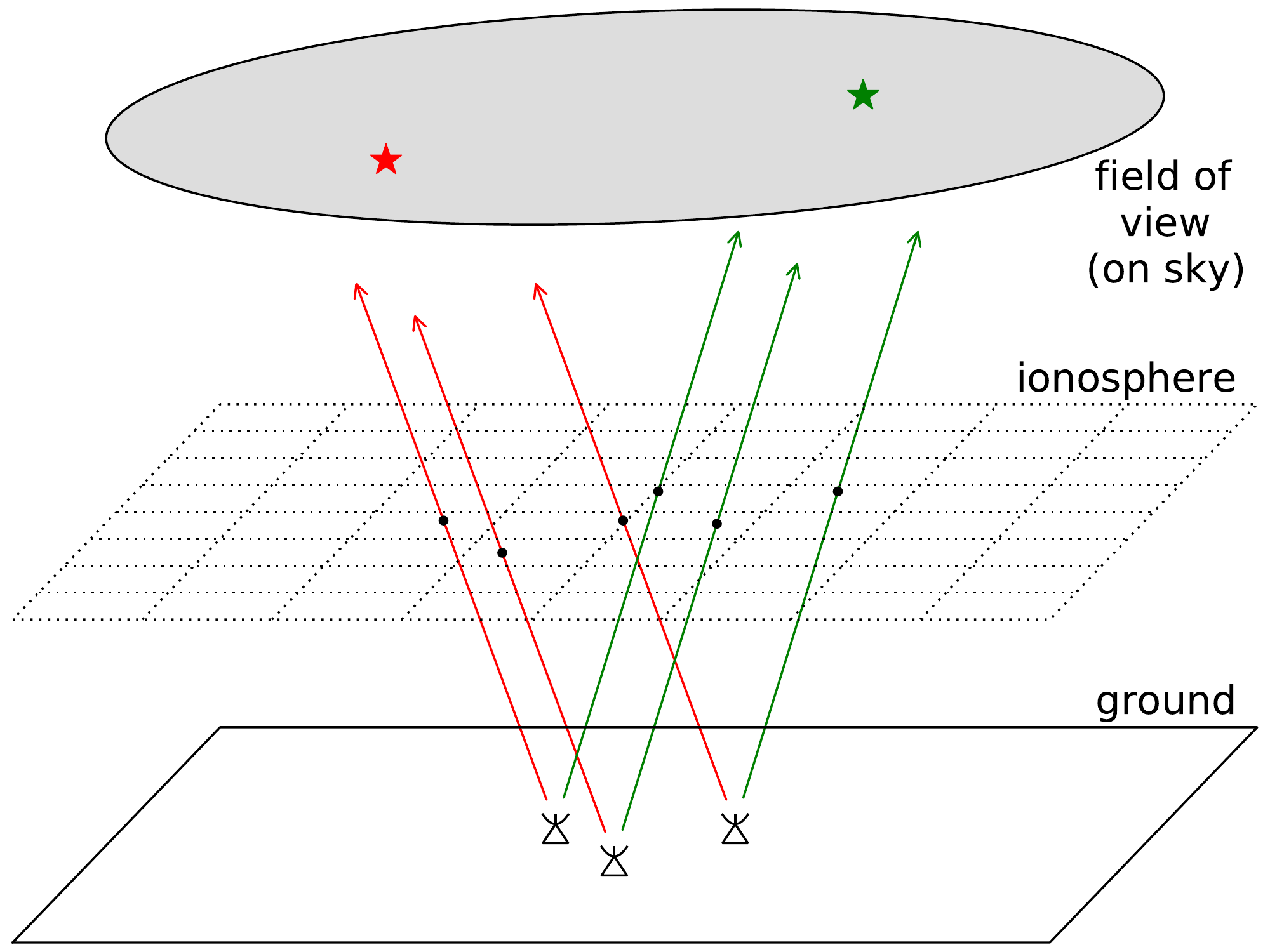}
\caption{{Illustration of \mbox{$n_{\rm ant} = 3$} antennas, part of a radio synthesis telescope, simultaneously observing \mbox{$n_{\rm src} = 2$} sources (in the far field).  This results in \mbox{$n_{\rm ant} \times n_{\rm src} = 6$} pierce points in the ionosphere (in the near field), which may be used for ionospheric calibration.}}
\label{fig:justinpiercepoints}
\end{center}
\end{figure}

{In low-frequency radio observations, there is typically substantial variation in the phase delay due to ionospheric structure across the field of view, as illustrated in} regimes{~3 and}~4 from {Figure~1 of} \citet{lonsdale05}.  {Ionospheric calibration can correct for this by using bright sources to form a model for the ionospheric structure, which then allows corrections to be applied for the ionospheric phase delay elsewhere in the field of view.}  In a typical radio observation at frequencies low enough to require ionospheric calibration, there are $n_{\rm ant} \gtrsim 20$ antennas observing \mbox{$n_{\rm src} \gtrsim 10$} radio sources bright enough to be used for ionospheric calibration in a single field of view \citep{vanweeren09,intema11,wykes14}{, which leads to \mbox{$n_{\rm ant} \times n_{\rm src} \gtrsim 200$} pierce points at which information can be obtained about the ionospheric electron content.  Figure~\ref{fig:justinpiercepoints} illustrates this for three antennas observing two sources, resulting in six pierce points; see also Figure~4 of \citet{intema09} for an example of ionospheric pierce points for a real observation.}

{SPAM (Source Peeling and Atmospheric Modeling; \citealt{intema09}), a widely-used method for performing ionospheric calibration, uses an iterative peeling procedure to determine the STEC along the line of sight between each source-antenna pair, relative to an arbitrary reference value.  In its original implementation, each line of sight is assumed to pierce the ionosphere at a single point as described by the thin-layer model, and the STEC at each of these pierce points is converted to an equivalent VTEC per Equation~\ref{eq:STEC}.  A power-law spectral density model is fitted to these points, allowing the VTEC to be interpolated across the footprint of the field of view on the ionosphere.  When imaging, these VTEC values are converted back to STEC via Equation~\ref{eq:STEC}, and the phase corrections calculated per Equation~\ref{eq:phiion}.  This procedure corrects for horizontal variation in the ionospheric electron content and, when repeated for successive time intervals, also corrects for temporal variation.}

{The use of the thin-layer model causes inaccuracy in ionospheric phase calibration through two different effects.  Firstly, lines of sight that nominally pass through the same point on the assumed thin-layer may sample different regions of the ionosphere above or below this altitude, which may have different electron content.  Secondly, even if horizontal variation in the ionospheric electron content is neglected, the curvature of the Earth leads to inaccuracy in the derived STEC, as discussed in Section~\ref{sec:thinlayermodel} and shown in Figure~\ref{fig:SlantTECtopplot2}.  In Section~\ref{sec:testingthinlayer} we investigate the magnitude of this latter error, as a measure of the range of validity of the thin-layer model, and of the circumstances under which it is necessary to use a more complex model such as those in later implementations of SPAM \citep{intema11}.}

%----------------------------------------------------------------------------------------------------------------------------------------------------------

\section{Testing the thin-layer model}
\label{sec:testingthinlayer}

\begin{figure}
\begin{center}
\includegraphics[width=\columnwidth]{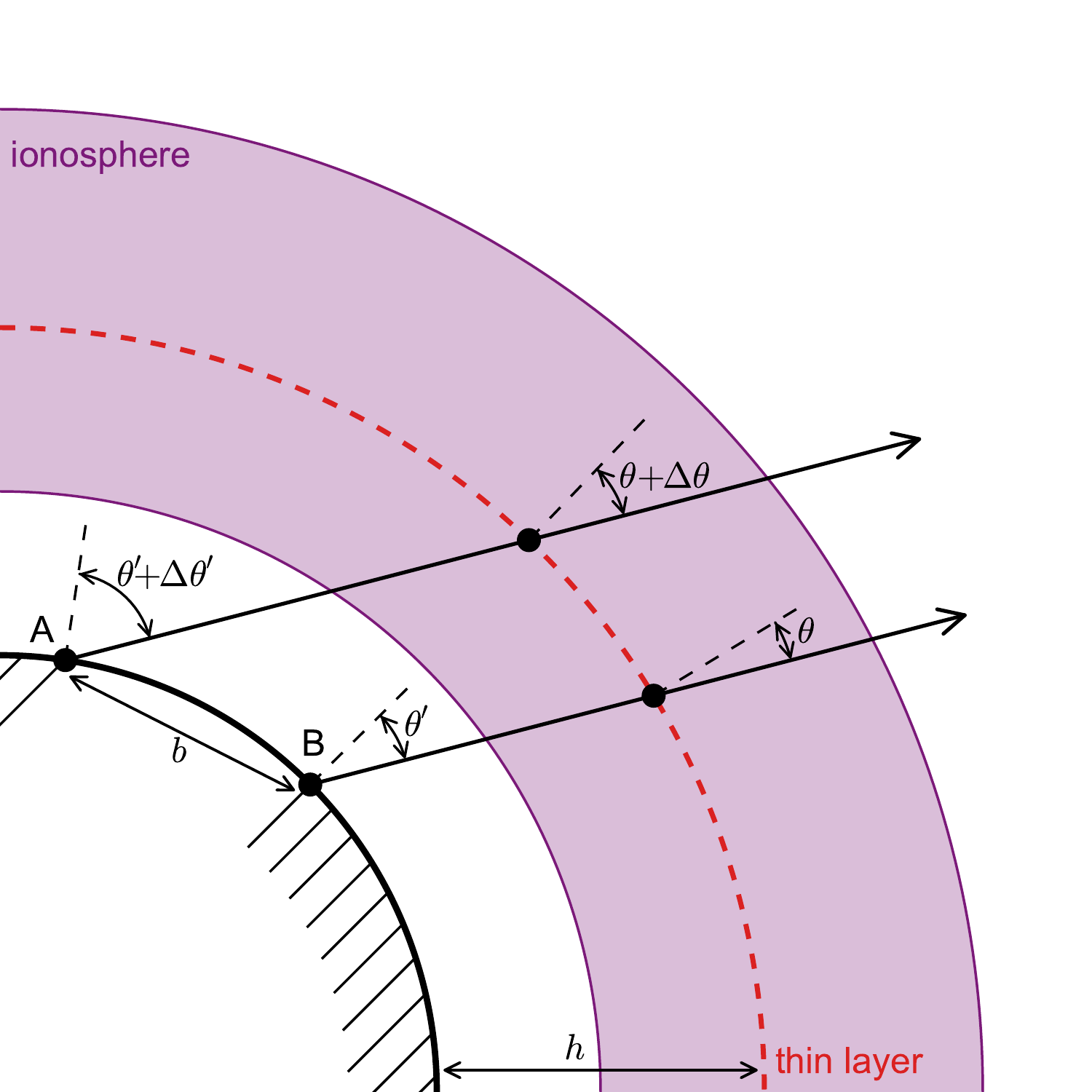}
\caption{{Illustration of two antennas (A and B), part of a radio synthesis telescope, observing a source through the ionosphere.  Due to the curvature of the Earth, the zenith angle $\theta'$ of the source from each antenna, and the consequent pierce angle $\theta$ at the thin-layer ionosphere, will differ, depending on the baseline length $b$.  The discrepancy $\Delta$STEC due to the thin-layer model will therefore differ between the two lines of sight, leading to a phase error $\Delta\phi_{\rm AB}$ across this baseline.}}
\label{fig:ionmod}
\end{center}
\end{figure}

{Consider two radio antennas, A and B, observing a source through the ionosphere as depicted in Figure~\ref{fig:ionmod}.  Let us assume that the ionosphere has been perfectly calibrated using the thin-layer model, and the VTEC is perfectly known at all points on the thin layer; for simplicity, we shall assume that this VTEC, and the electron density profile, are the same everywhere.  The derived STEC along the line of sight from each antenna will be subject to an error}
 \begin{equation}
  \Delta{\rm STEC} = {\rm STEC}_{\rm 2D}\ {-}\ {\rm STEC}_{\rm 3D}
 \end{equation}
{associated with the thin-layer model, as discussed in Section~\ref{sec:thinlayermodel}, which will lead to corresponding errors $\Delta\phi_{\rm A}$, $\Delta\phi_{\rm B}$ in the phase corrections.  If the antennas are closely spaced, these errors will be almost identical, and the error}
 \begin{equation}
  \Delta\phi_{{\rm AB}} = {\Delta}\phi_{\rm A} - {\Delta}\phi_{\rm B} \label{eq:deltaphi}
 \end{equation}
{in the phase correction on the baseline between these two antennas will be close to zero.  The larger the distance $b$ between the antennas, however, the more significant the error in the calibrated phase on this baseline.}

\begin{figure*}
\begin{center}
\includegraphics[width=\textwidth]{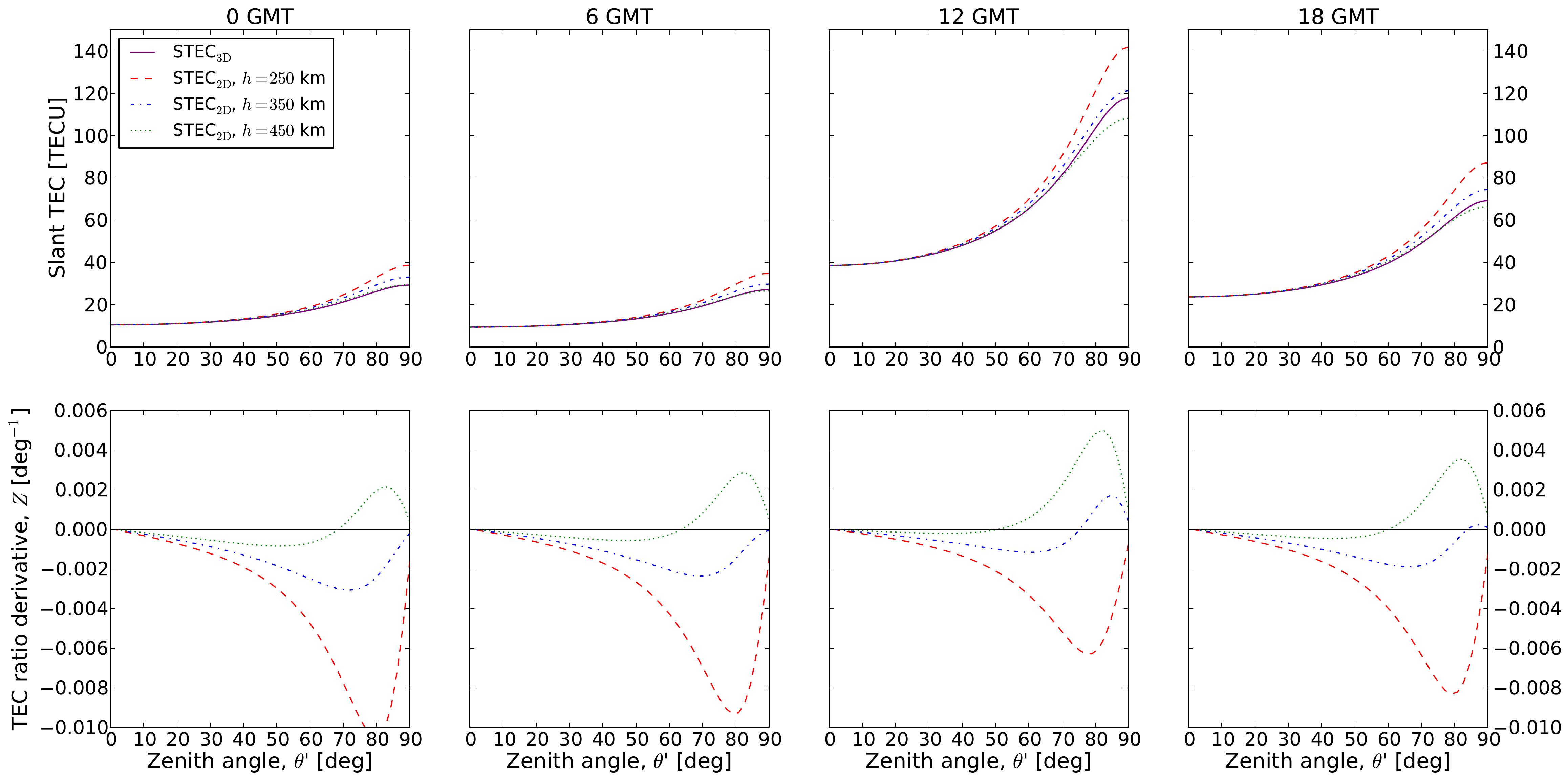}
\caption{Diurnal {variation of} STEC{$_{\rm 3D}$} and the thin-layer {model} STEC{$_{\rm 2D}$, and (lower panels) the} derivative $Z$ of the ratio between the {two.  The STEC shows the expected increase during the day due to solar forcing, while $Z$ varies significantly, but its magnitude is minimised at all times by a thin-layer altitude of \mbox{$h = 450$}~km, except at large zenith angles.}  Plots {are for the specified local times (GMT) on 17~February 2014 (local winter) for an ionospheric pierce point at $40^\circ$ geomagnetic latitude and $0^\circ$ geomagnetic longitude.}}
\label{fig:SlantTECplot_multipletimes}
\end{center}
\end{figure*}

\begin{figure*}
\begin{center}
\includegraphics[width=\textwidth]{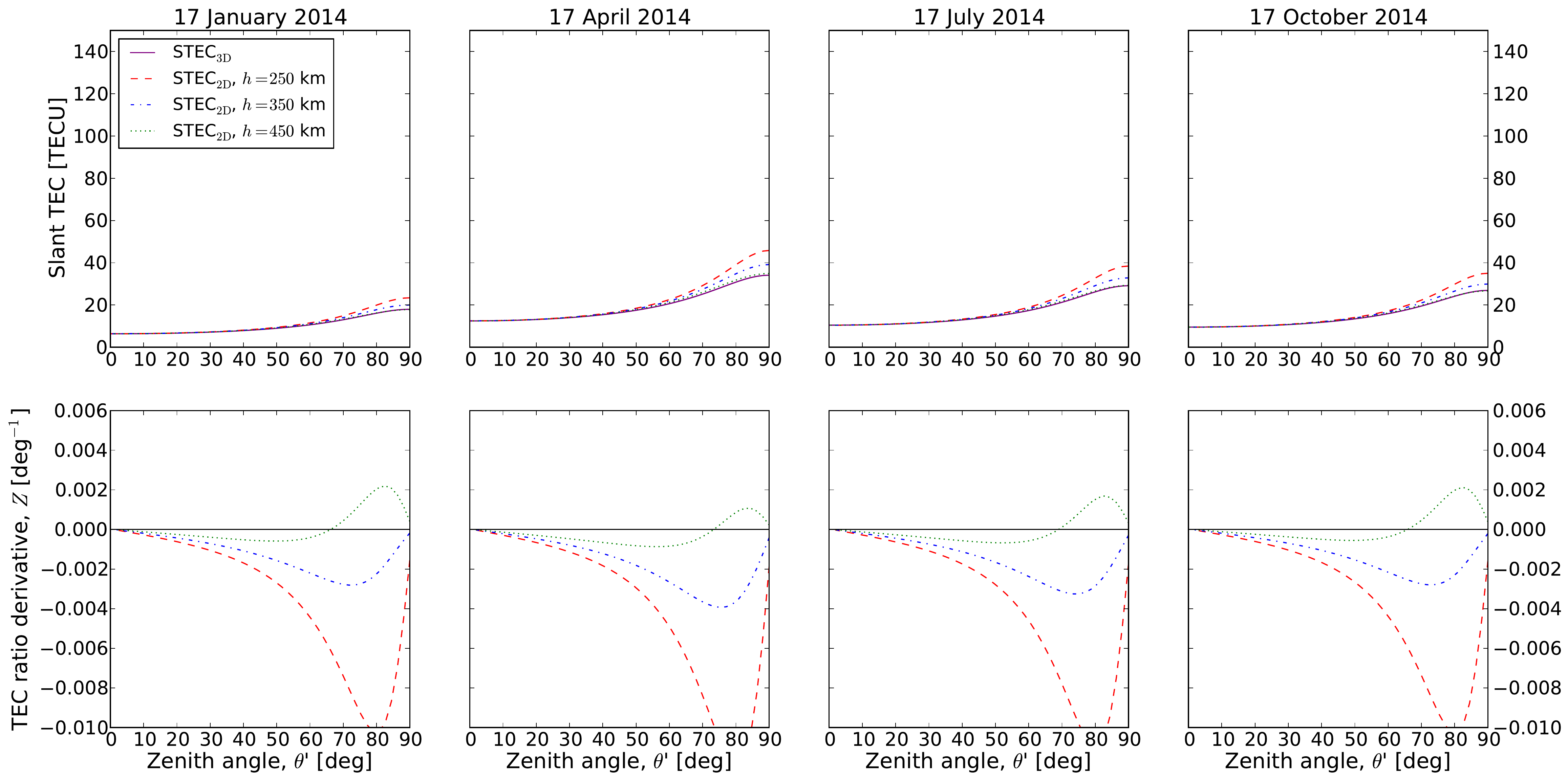}
\caption{Seasonal variation {of} STEC{$_{\rm 3D}$} and the thin-layer {model} STEC{$_{\rm 2D}$, and (lower panels) the} derivative $Z$ of the ratio between the {two. These plots are for {03:00} local time for an ionospheric pierce point at $40^\circ$ geomagnetic latitude and $0^\circ$ geomagnetic longitude. The STEC is maximised around April~2014 and July~2014 (local spring and summer, respectively), while $Z$ varies little throughout the year, its magnitude is always minimised by a thin-layer altitude of \mbox{$h = 450$}~km, except at large zenith angles. The variation with season differs depending on the latitude as it is affected by phenomena such as the mid-latitude winter anomaly (seen in the northern hemisphere) and the semiannual anomaly which is where the electron density at the F-layer peak height is greater at equinox than at solstice \citep{lee11}.}}
\label{fig:SlantTECplot_multipledates}
\end{center}
\end{figure*}

\begin{figure*}
\begin{center}
\includegraphics[width=\textwidth]{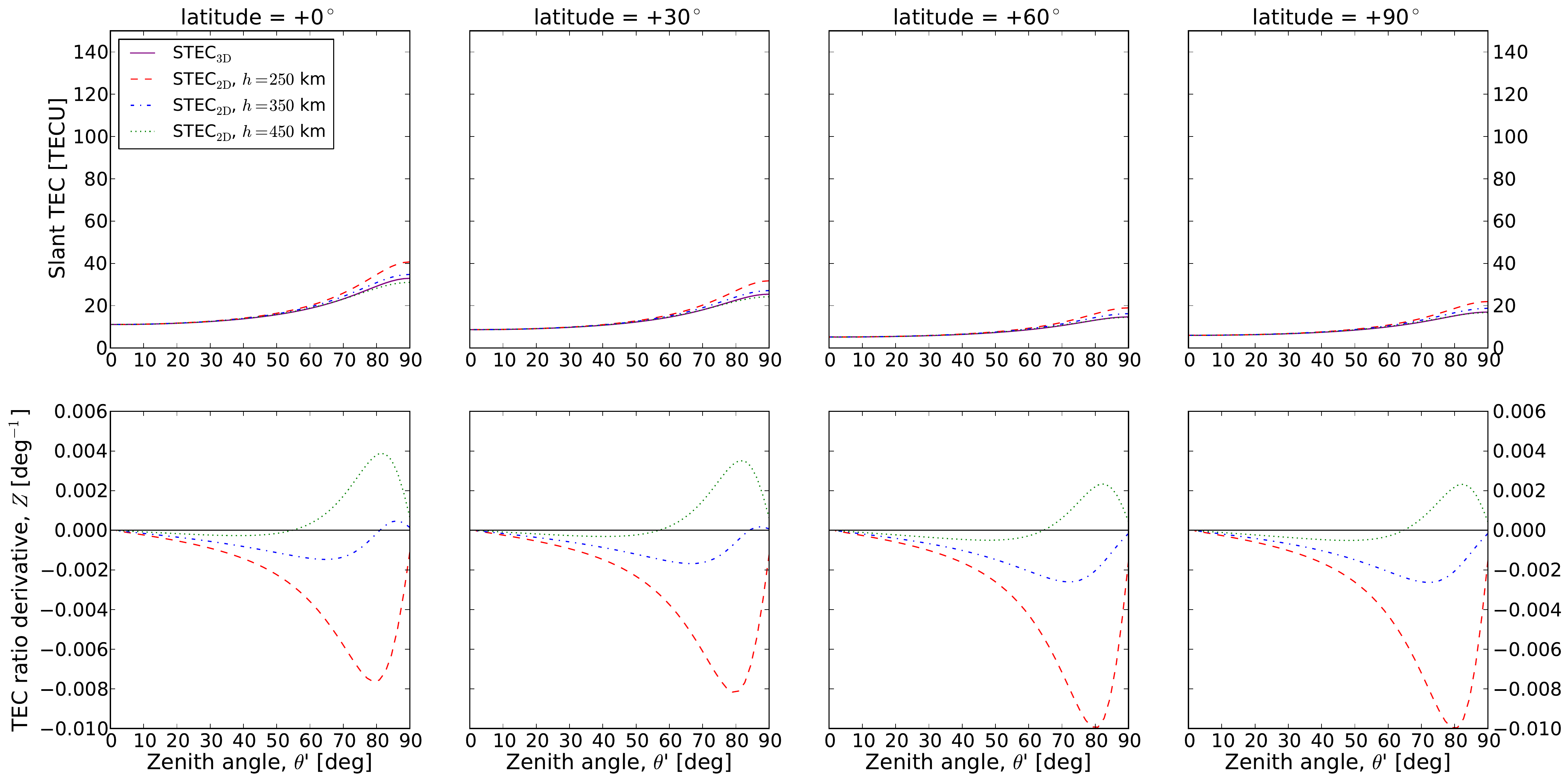}
\caption{{Variation} with latitude {of} STEC{$_{\rm 3D}$} and the thin-layer {model} STEC{$_{\rm 2D}$, and (lower panels) the} derivative $Z$ of the ratio between the {two.  The STEC is higher near the equator and lower near the poles. {The magnitude of $Z$ is minimised at all latitudes} by a thin-layer altitude of \mbox{$h = 450$}~km, except at large zenith angles. Plots are for {03:00} local time for an ionospheric pierce point at $0^\circ$ geomagnetic longitude on 17~February 2015 (local winter).}}
\label{fig:SlantTECplot_multiplelats}
\end{center}
\end{figure*}

\begin{figure}
\begin{center}
\includegraphics[width=\columnwidth]{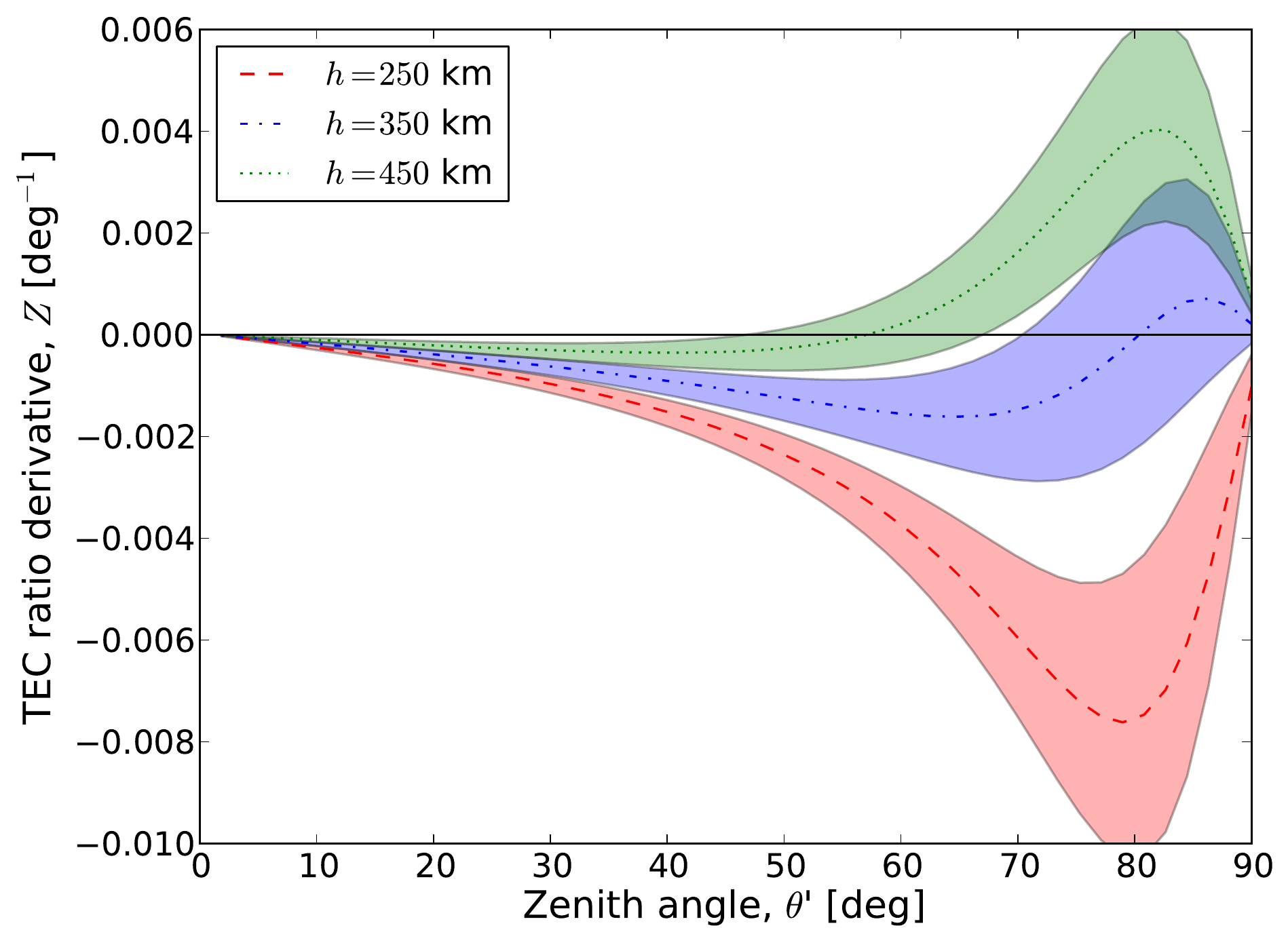}
\caption{Range {of possible values for the TEC} ratio derivative $Z${, reflecting} variation with time of day {(sampled at one-}hour {intervals), season (sampled} one day in {each} month{),} latitude {(from} $0^\circ$ {to} $90^\circ${), and} solar cycle {(i.e.\ over} the past 11~years{).  Lines show the median value for each thin-layer altitude $h$, and shaded areas denote the \mbox{$\pm 1\sigma$} range.  Values of $Z$ from this plot for the appropriate} zenith angle and thin-layer {altitude may be taken as typical inputs to Equation~\ref{eq:finalrelationship}.}}
\label{fig:SlantTECplot_everything}
\end{center}
\end{figure}

{The assumption here that VTEC is perfectly calibrated is not strictly accurate, because the calibration process is also based on phase measurements along lines of sight through the ionosphere, as described in Section~\ref{sec:ionocal}, and so is subject to the same discrepancy between the true STEC$_{\rm 3D}$ and the STEC$_{\rm 2D}$ values inferred for a thin-layer ionosphere.  However, it is the difference in this discrepancy from one antenna to another, due to differing zenith angles $\theta'$, that is responsible for the phase error $\Delta\phi_{\rm AB}$, and this difference is not corrected by the calibration procedure.  This is because the footprints on the ionosphere of the fields of view of different antennas typically overlap, as in Figure~\ref{fig:justinpiercepoints}, and so a point on the ionosphere will be calibrated on the basis of nearby pierce points of lines of sight from multiple antennas, with different zenith angles.}

{The phase error $\Delta\phi$ for each antenna will be related to $\Delta$STEC by Equation~\ref{eq:phiion}.  To find the phase error $\Delta\phi_{\rm AB}$ on the baseline, we need to find the difference between $\Delta$STEC for the two antennas, which we can approximate as
 \begin{equation}
  \Delta{\rm STEC}_{\rm AB} \approx \Delta\theta' \, \frac{d\Delta{\rm STEC}}{d\theta'} \label{eq:stecdiff}
 \end{equation}
in terms of the zenith angle $\theta'$, differing by $\Delta\theta'$ between the two antennas as shown in Figure~\ref{fig:ionmod}.  The first factor in this expression can be approximated as
 \begin{equation}
  \Delta\theta' \approx b / R_{\rm E} . \label{eq:earthcurve}
 \end{equation}
To find an expression for the second factor, we define the} TEC ratio derivative,
 \begin{equation}
  Z = \frac{d}{d\theta'} \frac{{\rm STEC}_{\rm 2D}}{{\rm STEC}_{\rm 3D}} \label{eq:tecrd} \\
 \end{equation}
{which we can approximate} as
 \begin{equation}
  Z \approx \frac{1}{{\rm STEC}_{\rm 3D}} \frac{d\Delta{\rm STEC}}{d\theta'}
 \end{equation}
{provided} that \mbox{$\Delta{\rm STEC}\ll{\rm STEC_{3D}}$}{, and further approximate as
 \begin{equation}
  Z \approx \frac{\cos\theta'}{\rm VTEC} \frac{d\Delta{\rm STEC}}{d\theta'} , \label{eq:tecrd_final}
 \end{equation}
provided that the line of sight is not close to the horizon.} 

{As the TEC ratio derivative $Z$ plays a key role in determining the phase error $\Delta\phi_{\rm AB}$, we investigate its behaviour across a range of parameters, using the same numerical approach as in Section~\ref{sec:thinlayermodel}.  Figures~\ref{fig:SlantTECplot_multipletimes}, \ref{fig:SlantTECplot_multipledates} and~\ref{fig:SlantTECplot_multiplelats} show respectively the variation with time of day, with season and with latitude.  In all cases, the strongest variation is with the zenith angle and the assumed altitude of the thin layer, with an altitude \mbox{$h = 450$}~km minimising the magnitude of $Z$ for zenith angles less than $55^\circ$.  The variation over all of these parameters is summarised in Figure~\ref{fig:SlantTECplot_everything}, which provides a range of typical values for $Z$.} 

{Substituting Equations~\ref{eq:earthcurve} and~\ref{eq:tecrd_final} into Equation~\ref{eq:stecdiff}, and converting from STEC to phase with Equation~\ref{eq:phiion}, we find the phase error associated with the thin-layer model, on the baseline between two antennas, to be}
 \begin{align}
  \Delta \phi_{\rm {AB}} &\approx 0.44^\circ \times
   \left(\frac{\nu}{\rm{100~MHz}}\right)^{-1}
   \left(\frac{b}{\rm{km}}\right)
   \nonumber \\ & \times
   \left(\frac{{\rm VTEC}}{\rm{10~TECU}}\right)
   \left(\frac{Z}{{\rm 0.001~deg}^{-1}}\right)
   \frac{1}{\cos\theta'} .
  \label{eq:finalrelationship}
 \end{align}
{For comparison, { \citet{perley99} states that a radio image typically has a maximum dynamic range of 20,000:1 which corresponds to a phase error on a baseline of}
\begin{equation}
\phi\approx\frac{N}{D}
\label{eq:perleydynamic}
\end{equation}
{where $N$ is the number of antennas and $D$ is the dynamic range. Residual ionospheric phase errors will cause a systematic error in the phase measured on each antenna, however due to the largely-random distribution of antennas in an array this can be treated as random error in a similar way to how the phase errors resulting from inaccuracy in baseline determination are treated by \citet{perley99}. If $\Delta\phi_{\rm AB}$ exceeds the value given by Equation~\ref{eq:perleydynamic}}{, the use of the thin-layer model will degrade the dynamic range of an image.}

\begin{table*}
\setlength{\tabcolsep}{2pt}
\caption{{Typical phase error resulting from the thin-layer {model} for {five} radio synthesis telescopes: the GMRT, the MWA, LOFAR {LBA} {and HBA} (Dutch stations, core and Superterp) {the {VLA} (A and D configurations) and SKA-low}. The table displays the frequency, latitude, maximum baseline and magnitude of the TEC ratio derivative, $Z$, assuming a zenith angle of $\theta' = 45^{\circ}$ for each telescope. {The lowest frequencies are considered for each telescope, providing a worst-case scenario for observers.} The phase error is calculated for each telescope {for typical mid-latitude daytime (10~TECU) and nighttime (4~TECU) VTEC \citep{verkhoglyadova13}}, assuming a pointing azimuth of either due east or due west (i.e. fixed latitude) and a thin-layer altitude of 450~km. The VTEC can vary by a factor of 10 { from the typical mid-latitude daytime VTEC} and as the VTEC varies, it will cause a proportional change in the phase error. { Assuming a dynamic range of 20,000:1, the limiting phase error is calculated for each telescope for comparison}.}}
\label{tab:telescopecheck} 
\begin{center}
\begin{tabular}{lr@{.}lr@{.}lr@{.}lr@{.}lr@{.}lr@{.}lrr@{.}l}
\hline
Telescope &  \multicolumn{2}{l}{Frequency} & \multicolumn{2}{l}{Latitude} &  \multicolumn{2}{l}{Baseline} &  \multicolumn{2}{l}{$|Z|$} &  \multicolumn{2}{l}{{Actual phase error}} &  \multicolumn{2}{l}{Actual phase error} & {Number of} & \multicolumn{2}{l}{{Max allowed}}  \\
 & \multicolumn{2}{l}{[MHz]} &  \multicolumn{2}{l}{[deg]} &  \multicolumn{2}{l}{[km]} & \multicolumn{2}{l}{[deg$^{-1}$]} & \multicolumn{2}{l}{{at 4 TECU [$^{\circ}$]}} & \multicolumn{2}{l}{at 10 TECU [$^{\circ}$]} & {antennas} &  \multicolumn{2}{l}{{phase error [$^{\circ}$]}} \\ 
\hline
GMRT & 153&0 & $+$19&10 & 25&0 & 0&00020 & 0&80 & 2&01 & 30 & 0&09  \\
LOFAR {LBA} (Superterp) & 30&0 & $+$52&91 & 0&24 & 0&00015 & 0&03 & 0&07 & 6 & 0&02  \\
LOFAR {LBA} (core) & 30&0 & $+$52&91 & 3&5 & 0&00015 & 0&43 &1&08 & 24 & 0&07 \\
LOFAR {LBA} (Dutch stations) & 30&0 & $+$52&91 & 121&0 & 0&00015 & 14&89 & 37&22 & 16 & 0&11 \\
{LOFAR {HBA} (Superterp)} & 110&0 & $+$52&91 & 0&24 & 0&00015 & 0&01 & 0&02 & 6 & 0&02 \\
{LOFAR {HBA} (core)} & 110&0 & $+$52&91 & 3&5 & 0&00015 & 0&12 & 0&29 & 24 & 0&07 \\
{LOFAR {HBA} (Dutch stations)} & 110&0 & $+$52&91 & 121&0 & 0&00015 & 4&06 & 10&15 & 16 & 0&11  \\
MWA & 80&0 &  $-$26&70 &  2&86 & 0&00005 & 0&04 & 0&11 & 128 & 0&37  \\
{{VLA}} (A configuration) & 58&0 & $+$34&08 & 36&4 & 0&00010 & 1&54 & 3&86 & 27 & 0&08 \\
{{VLA}} (D configuration) & 58&0 & $+$34&08 & 1&03 & 0&00010 & 0&04 & 0&11 & 27 & 0&08 \\
{SKA-low} & 50&0 & $-$26&80 & 65&0 & 0&00005 & 1&60 & 4&00 & 500 & 1&43 \\
\hline
\end{tabular}
\end{center}
\end{table*}

{In Table~\ref{tab:telescopecheck} we calculate the phase calibration precision required to achieve this dynamic range using Equation~\ref{eq:perleydynamic} for the radio synthesis telescopes mentioned in Section~\ref{sec:intro} (LOFAR, GMRT, {{VLA}, MWA and SKA-low}) and, as examples, calculate the {typical values} of the} phase error for the longest baselines of {these} radio {synthesis} telescopes. {For most of these {arrays} (the {VLA}, the GMRT, LOFAR and the SKA-low), the use of the thin-layer {model} will be a limiting factor at the {listed} frequencies assuming a typical mid-latitude, daytime VTEC of 10~TECU \citep{verkhoglyadova13}, {but} for {the more compact MWA, the use of the thin-layer {model} is not {a} limiting factor unless the ionosphere is extremely active {and the VTEC is correspondingly high}}. If a typical mid-latitude, nighttime VTEC of 4~TECU \citep{verkhoglyadova13} is instead assumed, the use of the thin-layer {model} is not {a} limiting factor for the LOFAR HBA (Superterp) and the {VLA} in D-configuration as well as for the MWA}.

%----------------------------------------------------------------------------------------------------------------------------------------------------------

\section{Conclusion}
\label{sec:conc}

We have {numerically} investigated the {phase error that results from using} the thin-layer {model} {in ionospheric phase calibration for} a radio {synthesis} telescope{, independent of horizontal variation in the ionospheric electron content, using a realistic vertical electron density profile.  We find that an assumed} thin-layer altitude of \mbox{${\sim} 450$}~km {generally minimises this particular error.  Our final result allows} observers {to determine the circumstances under which} the thin-layer {model} can {reasonably} be applied to radio observations{:} Equation~\ref{eq:finalrelationship}{, together with typical values for the TEC ratio derivative $Z$ from} Figure~\ref{fig:SlantTECplot_everything}, {provides a value for the phase error resulting from the use of the} thin-layer {model}. {We find that this phase error limits the imaging fidelity for long-baseline arrays such as the {{VLA}, the GMRT, and LOFAR LBA array and HBA array, as well as the future SKA-low}, as shown in Table~\ref{tab:telescopecheck}.}

%----------------------------------------------------------------------------------------------------------------------------------------------------------

%\newpage

\bibliographystyle{mnras}
\bibliography{mybib} % if your bibtex file is called example.bib

% Don't change these lines
\bsp	% typesetting comment
\label{lastpage}
\end{document}